\documentclass[aps,pre,twocolumn,footinbib,superscriptaddress]{revtex4-1}

\usepackage{amsmath}
\usepackage{array}
\usepackage{graphicx}
\usepackage{dcolumn}
\usepackage{bm}
\usepackage{color}
\usepackage[unicode]{hyperref}
\hypersetup{
   unicode=true,          
   plainpages=false,
   colorlinks=true,       
   linkcolor=red,          
   citecolor=blue,        
   filecolor=green,      
}

\newcolumntype {s}[1]{@{\hspace{#1}}} 
\newcolumntype {R}{>{$}r<{$}}         
\newcolumntype {C}{>{$}c<{$}}         
\newcolumntype {L}{>{$}l<{$}}         
\newcolumntype {f}{@{\extracolsep\fill}}  

\graphicspath{{.}{./Figs/}}

\begin{document}

\title{\textbf{\textit{Ex-ante\/}} measure of patent quality reveals
intrinsic fitness for citation-network growth}

\author{K.~W. Higham}
\altaffiliation[Present address: ]{College of Management of Technology,
EPFL, Odyssea, Station 5, 1015 Lausanne, Switzerland}
\affiliation{Te P{\=u}naha Matatini, School of Chemical and Physical
Sciences, Victoria University of Wellington, PO Box 600, Wellington 6140,
New Zealand}

\author{M. Governale}
\affiliation{Te P{\=u}naha Matatini, School of Chemical and Physical
Sciences, Victoria University of Wellington, PO Box 600, Wellington 6140,
New Zealand}

\author{A.~B. Jaffe}
\affiliation{Te P{\=u}naha Matatini, Motu Economic and Public Policy
Research, PO Box 24390, Wellington 6142, New Zealand}
\affiliation{MIT Sloan School of Management, 100 Main Street, Cambridge,
MA 02142}
\affiliation{Brandeis University, 415 South Street, Waltham, MA 02453}
\affiliation{QUT Business School, Queensland University of Technology,
Brisbane, QLD 4001, Australia}

\author{U. Z\"ulicke}
\email{uli.zuelicke@vuw.ac.nz}
\affiliation{Te P{\=u}naha Matatini, School of Chemical and Physical
Sciences, Victoria University of Wellington, PO Box 600, Wellington 6140,
New Zealand}

\begin{abstract}
We have constructed a fitness parameter, characterizing the intrinsic
attractiveness for patents to be cited, from attributes of the associated
inventions known at the time a patent is granted. This exogenously
obtained fitness is shown to determine the temporal growth of the
citation network in conjunction with mechanisms of preferential
attachment and obsolescence-induced ageing that operate without reference
to characteristics of individual patents. Our study opens a window on
understanding quantitatively the interplay of the
\textit{rich-gets-richer} and \textit{fit-gets-richer} paradigms that
have been suggested to govern the growth dynamics of real-world complex
networks.
\end{abstract}

\date{\today}

\maketitle

A wide variety of social and economic processes evolve such that success
or popularity appear to be self-reinforcing. Significant attention has
been given to trying to distinguish the extent to which the apparently
self-reinforcing behavior of popularity is purely a property of the
dynamic system, versus being generated by intrinsic heterogeneity that
allows inherently better agents or products to persistently
succeed~\cite{sou07,kon08,pap12,wan13,rij14,sin16,zen17,for18,mar18}.
With our increasingly data-rich world enabling more effective ways for
measuring quality as well as popularity, this question can now be
explored more deeply and in a greater variety of fields~\cite{ves12}.
Here we provide an answer within the context of technological innovation,
where an accepted measure of popularity is the intensity with which
patents accumulate citations~\cite{jaf17}. Our construction of a
technology-dependent single quality score for individual patents from a
broad range of patent-quality measures that are exogenous to citations
and available at the time of grant is shown to quantify the innate
attractiveness of patents to be cited in the future. The ability to
account for inherent quality as a driver of citation dynamics enables
better observation of other important influences, including the time
scale for knowledge obsolescence~\cite{hig17}.

Empirically, the average rate $\bar\lambda$ at which the number of
citations $k$ accrued by patents~\cite{hal02} increases over time $t$ is
observed~\cite{csa07,val07,hig17} to follow an
aging-tempered~\cite{dor00} preferential-attachment-type~\cite{pri76,
bar99,kra01} growth model $\bar\lambda = A(t)\, f(k)$. The asymptotic
form $f(k)\sim k^\alpha$ for large $k$ with $\alpha > 0$ embodies a
\emph{rich-gets-richer\/} feedback loop whereby more highly cited patents
are more likely to gain future citations. Similar behavior is exhibited
by the citation dynamics of scientific articles, e.g., those published in
the journals of the American Physical Society~\cite{red05,gol12,hig17b}.
However, the special purpose and associated legal ramifications of
citations in patents~\cite{hal02,jaf17} enforce a greater degree of
caution in citing behavior than is commonly practiced for scientific
articles, making patent citations particularly suitable for investigating
the relationship between popularity and
quality~\footnote{Qualitative~\cite{mey00} and quantitative~\cite{clo15}
studies have shown that patent citations are less likely to be irrelevant
or superfluous. Also, unlike scientific articles, patent publications
contain a highly regulated set of metadata suitable for constructing
universal intrinsic-quality indicators.}.

On a phenomenological level, purely preferential-attachment-based models
seem to be able to successfully describe the dynamics of how patents
receive forward citations. However, they are at odds with the general
expectation~\cite{mar07} that patents are intrinsically heterogeneous in
quality, and that citing behavior will, at least in part, be influenced
by this intrinsic heterogeneity across the patent population. In
theoretical models of network growth, node heterogeneity has been
introduced by a fitness, or attractiveness, variable $\eta$ with
distribution $\rho(\eta)$ so that an individual node $i$ having $k_i$
links to other nodes at time $t$ after its creation gains new links with
a rate~\cite{bia01}
\begin{equation}\label{eq:fitRate}
\lambda_i = \eta_i\, \tilde A(t)\, \tilde f(k_i) \quad .
\end{equation}
In principle, a fitness variable~\footnote{Without loss of generality, we
assume fitness $\eta$ to be normalized such that its mean satisfies
$\mu_\eta=1$.}  can represent any quantifiable heterogeneous property, or
a number of such properties, exhibited by individual nodes~\cite{fer12},
and it can even be designed to depend on the properties of the linking
node~\cite{pap12,fer12}. The interplay between fitness and preferential
attachment has been studied in detail theoretically~\cite{pap12,gol18,
pha16}, and statistical analyses have been applied to estimate fitness
endogenously in real-world networks~\cite{new07,kon08,wan13,pha16,ron18}.
Our current work goes an important step further by determining fitness
for individual patents in terms of citation-independent quality measures.
Obtaining fitness deterministically and exogenously enables us to
conclusively separate its effects from those due to preferential
attachment and obsolescence-induced ageing~\footnote{A rate of the form
given in Eq.~(\ref{eq:fitRate}) covers the extreme cases of growth by
pure preferential attachment~\cite{bar99} [by letting $\rho(\eta)\to
\delta(\eta - 1)$, where $\delta (\cdot)$ is the Dirac-$\delta$ function]
or pure fitness~\cite{cal02,gar04,tac12} [$\tilde f(k)\to
\mathrm{constant}$].}. These advances pave the way for broader studies of
knowledge diffusion and could inform the design of meaningful impact
measures for technological innovation.

How to determine the quality and/or value of innovations from observable
attributes of the associated patents is a difficult question that has
been studied extensively~\cite{lan04,gam08,zee11,kog17,ras18}. As there
are various potential dimensions of usefulness for an invention
(including, but not limited to, technological, economic, and
strategic/legal), the number of suggested plausible quality-indicator
variables has proliferated. We base our construction of a single
intrinsic quality score for patents on the $N_v=13$ variables $v^{(a)}$
given in Table~\ref{tab:qvars}. Their values $v^{(a)}_i$ for a particular
patent $i$ are determined by the time of grant and become part of its
retrievable official record. These quality measures are also available
for all patents, not just a subset such as those assigned to publicly
traded companies~\cite{kog17}. However, before we can use the values
$v^{(a)}_i$ to estimate the intrinsic fitness $\eta_i$ for a patent to
attract citations via Eq.~(\ref{eq:fitRate}), three major issues need to
be addressed: (i)~minimizing cross-correlations (the variables $v^{(a)}$
are not necessarily mutually independent quality indicators),
(ii)~determining relative weighting (multiple uncorrelated quality
measures will differ in the magnitude of their influence on citation
growth), and (iii)~achieving distributional fidelity (the model for
fitness constructed in terms of the variables $v^{(a)}$ must reproduce
salient properties of the empirical distribution of fitness). We now
describe in turn how each of these challenges is addressed.

\begin{figure}[t]
\includegraphics[height=3.7cm]{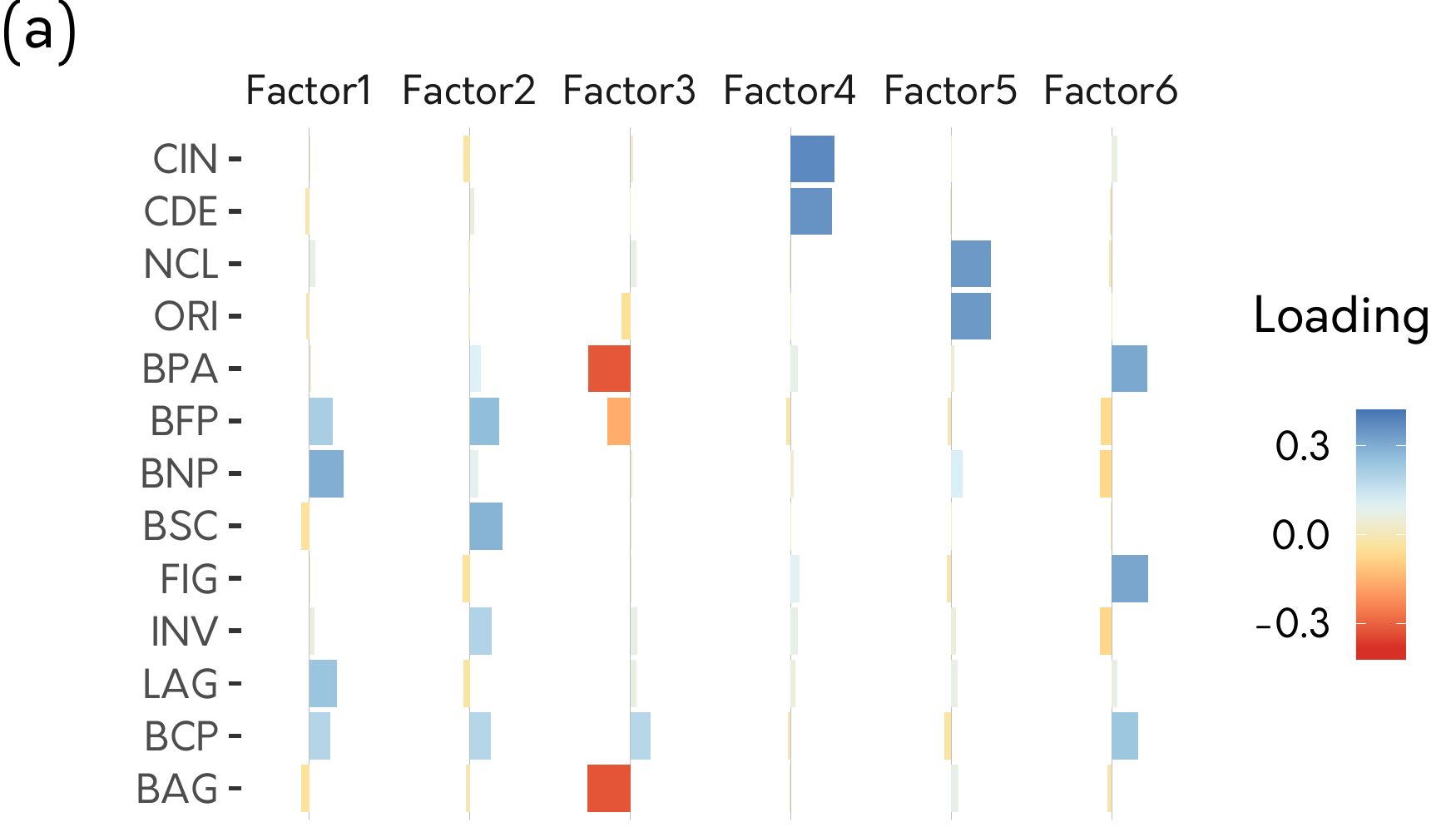}\hfill
\includegraphics[height=3.7cm]{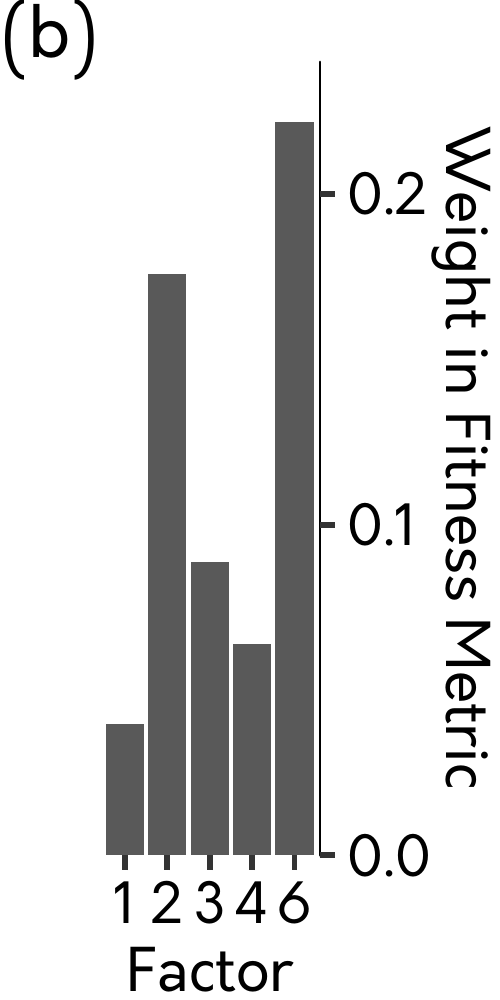}
\caption{\label{fig:EFAres}
Distilling a fitness for citation-network growth from the $N_v=13$
quality indicators in Table~\ref{tab:qvars} for the cohort of USPTO
patents from CPC Section A with grant dates during 1999-2001. Factor
analysis yields $N_u=6$ factors $\bar u^{(b)}$ representing minimally
correlated quality-indicator variances across the patent cohort.
Panel~(a) illustrates the loadings $L_a^{(b)}$ that relate factor scores
$\bar u^{(b)}_i$ for individual patents to the rescaled logs of raw
quality-indicator values [cf.\ Eq.~(\ref{eq:logVars})] via $\bar
u^{(b)}_i =\sum_{a=1}^{N_v} L^{(b)}_a \, \bar v^{(a)}_i$. Panel (b)
shows the relative weights $w_b$ for each factor in the expression
Eq.~(\ref{eq:fitDef}) for the fitness variable, which are obtained by a
forward-stepwise regression to the distribution of citations accrued by
the time of grant by patents from a small training set. Here factor $\bar
u^{(5)}$ was eliminated according to the Bayes Information Criterion.}
\end{figure}

\textit{(i)~Minimizing cross-correlations.\/} As it is not \textit{a
priori\/} clear which type of quality is measured by each individual
variable and/or how much overlap exists between the quality measures
provided by different variables, an exploratory factor
analysis~\cite{mul10} is performed~\cite{noteSM}. This process yields
$N_u < N_v$ uncorrelated factors $\bar u^{(b)}$ that are linear functions
of the $N_v$ normalized variables~\footnote{We use $\mu_q$ and $\sigma_q$
to indicate the mean and variance, respectively, of any randomly
distributed quantity $q$.}
\begin{table}[t]
\centering
\caption{\label{tab:qvars}
Variables $v^{(a)}$ employed as raw measures of intrinsic patent quality.
Short labels given in brackets are introduced for easy reference.}
\begin{tabular}{|l|l|}\hline
backward citations to patents & independent claims $-\, 1$ \\
(BPA)  & (CIN) \\ \hline
backward self-citations (BSC) & dependent claims (CDE) \\ \hline
backward citations to foreign & inventor team size $-\, 1$ \\
patents (BFP) & (INV)  \\ \hline
backward citations to non-patent & class membership $-\, 1$ \\ 
literature (BNP) & (NCL) \\ \hline
backward-citations' pedigree~\cite{noteSM} & average age of backward \\ 
(BCP) & citations (BAG) \\ \hline
originality~\cite{noteSM} (ORI) & grant lag (LAG) \\ \hline
& number of figures (FIG) \\ \hline
\end{tabular}
\end{table}
\begin{equation}\label{eq:logVars}
\bar v^{(a)} = \left\{ \begin{array}{cl}
\frac{\ln(1+v^{(a)})-\mu_{\ln(1+v^{(a)})}}{\sigma_{\ln(1+v^{(a)})}} &
\mbox{all variables except ORI,} \\[0.3cm]
\frac{v^{(a)} - \mu_{v^{(a)}}}{\sigma_{v^{(a)}}} & \mbox{for ORI.}
\end{array} \right.
\end{equation}
The definition of $\bar v^{(a)}$ according to Eq.~(\ref{eq:logVars}) is
motivated by the observation that, with the exception of the normally
distributed originality, the raw indicator variables $v^{(a)}$ are
approximately log-normally distributed over the patent cohorts considered
in this work. It furthermore ensures that the $\bar v^{(a)}$ have zero
mean and unit variance, thus eliminating the arbitrary units used to
measure each $v^{(a)}$. The factors $\bar u^{(b)}$ also have zero mean
and unit variance, and their values for individual patents are determined
via the matrix of loadings $L^{(b)}_a$ that is obtained from the factor
analysis: $\bar u^{(b)}_i = \sum_{a=1}^{N_v} L^{(b)}_a\,\bar v^{(a)}_i$.
Figure~\ref{fig:EFAres}(a) illustrates the factors and their loadings
obtained for the cohort of patents granted by the United States Patent
and Trademark Office (USPTO) between 1 January 1999 and 31 December 2001
and classified under Section A (Human Necessities) of the Cooperative
Patent Classification (CPC) system. We have repeated this analysis for
another six cohorts of USPTO patents that are granted within the same
period but classified under different CPC Sections: B (Performing
Operations; Transporting), C (Chemistry; Metallurgy), F (Mechanical
Engineering; Lighting etc.), G (Physics), H (Electricity), and Y
(General), respectively~\cite{noteSM}.

\begin{figure*}[t]
\includegraphics[width=0.245\textwidth]{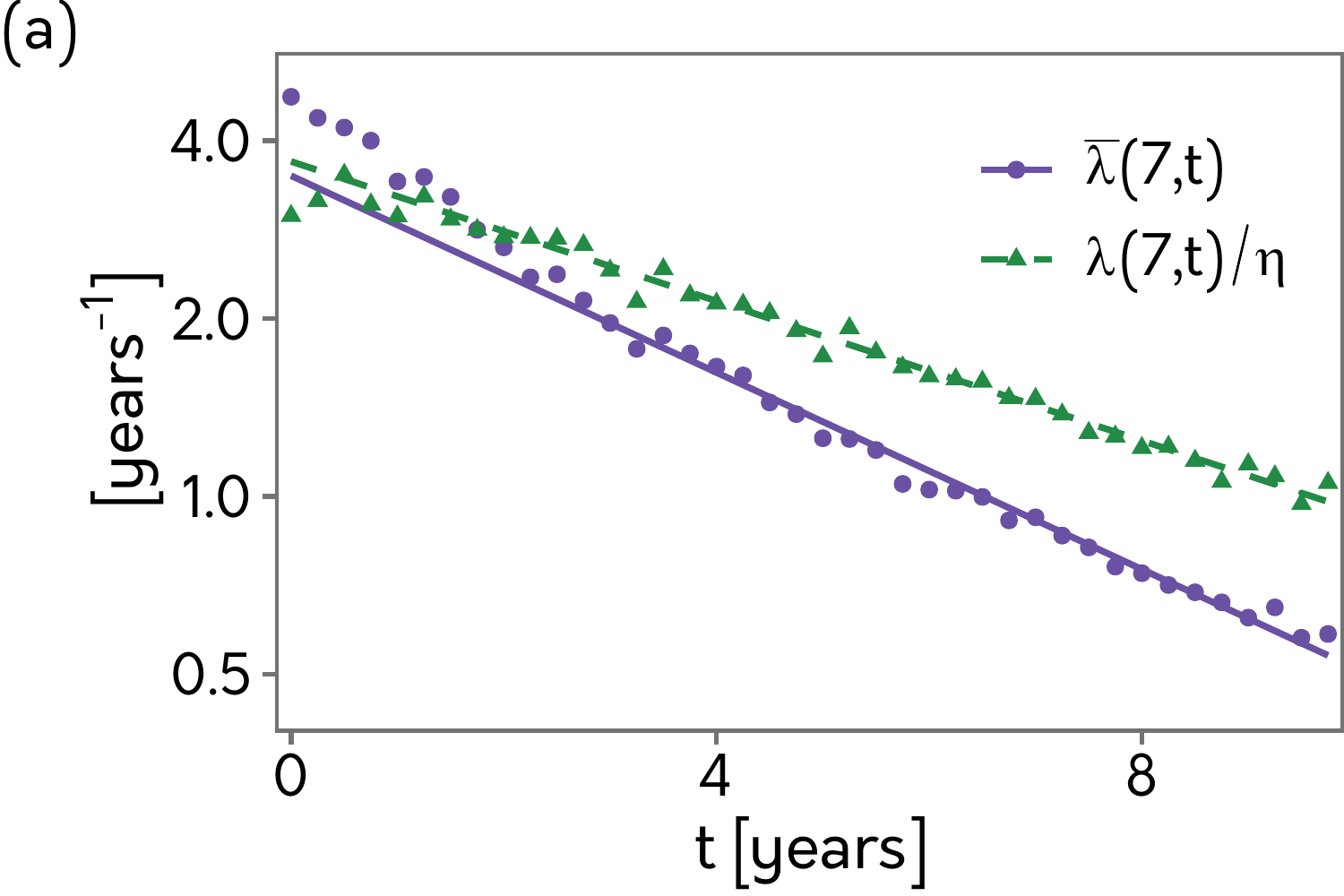}\hfill
\includegraphics[width=0.245\textwidth]{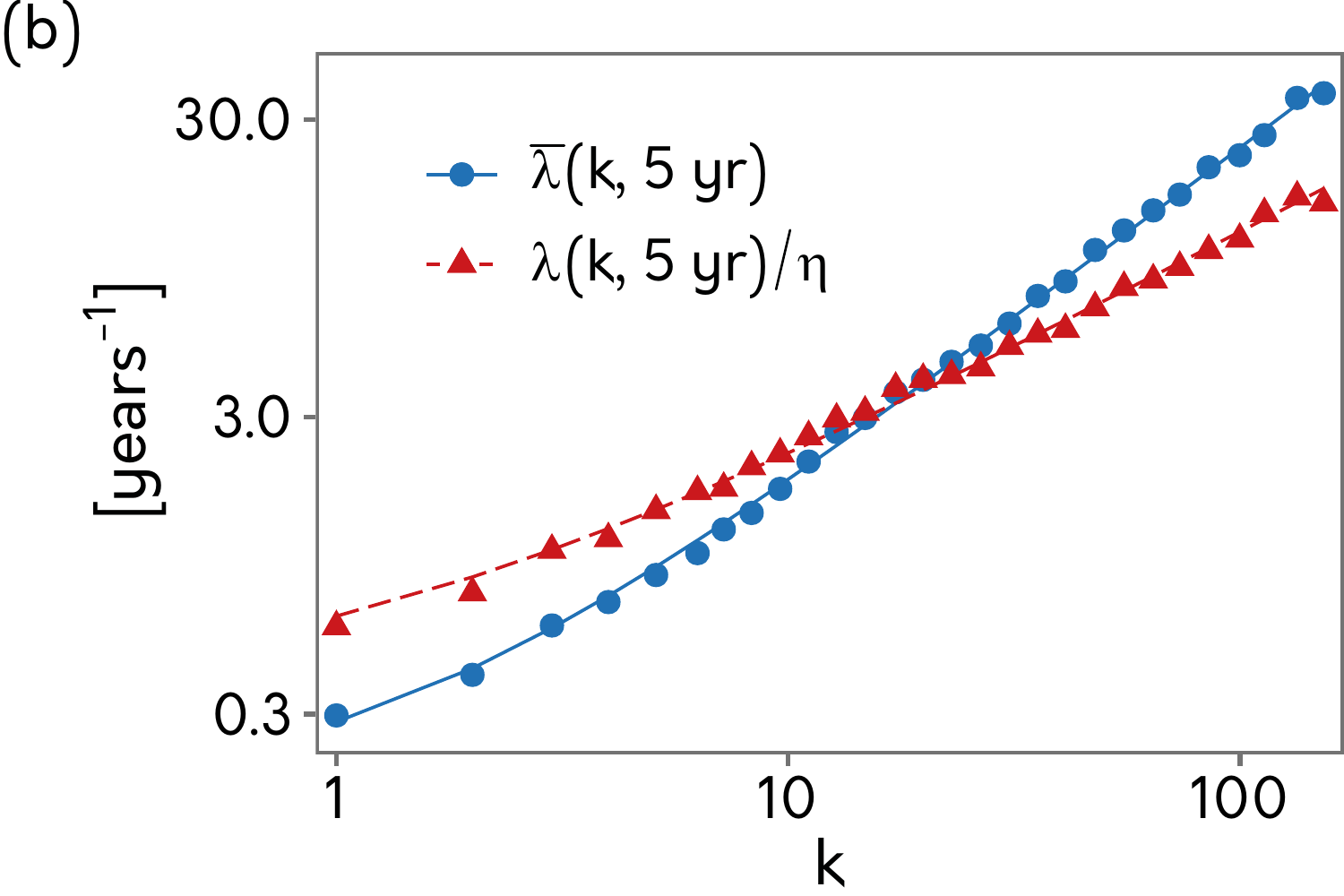}\hfill
\includegraphics[width=0.245\textwidth]{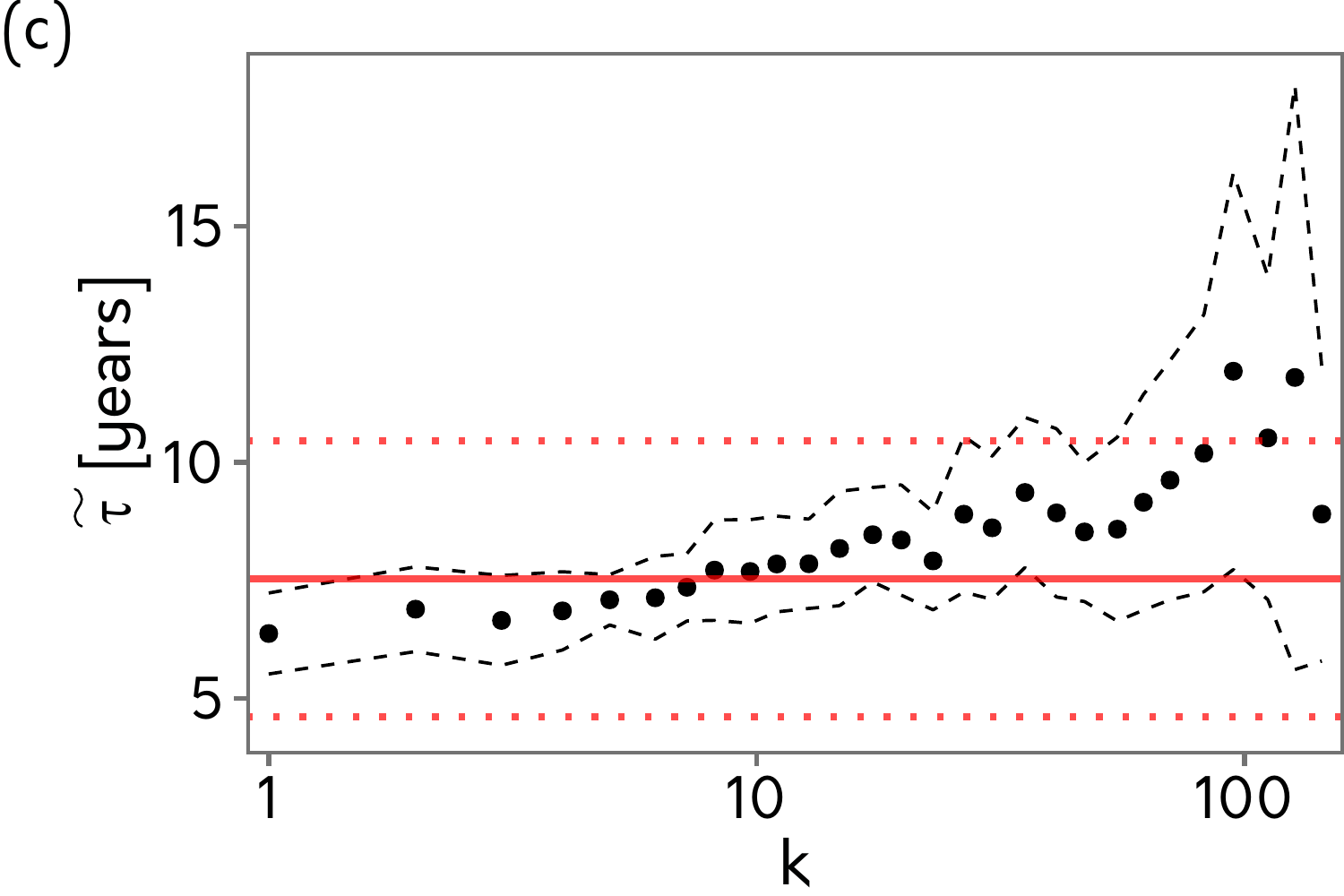}\hfill
\includegraphics[width=0.245\textwidth]{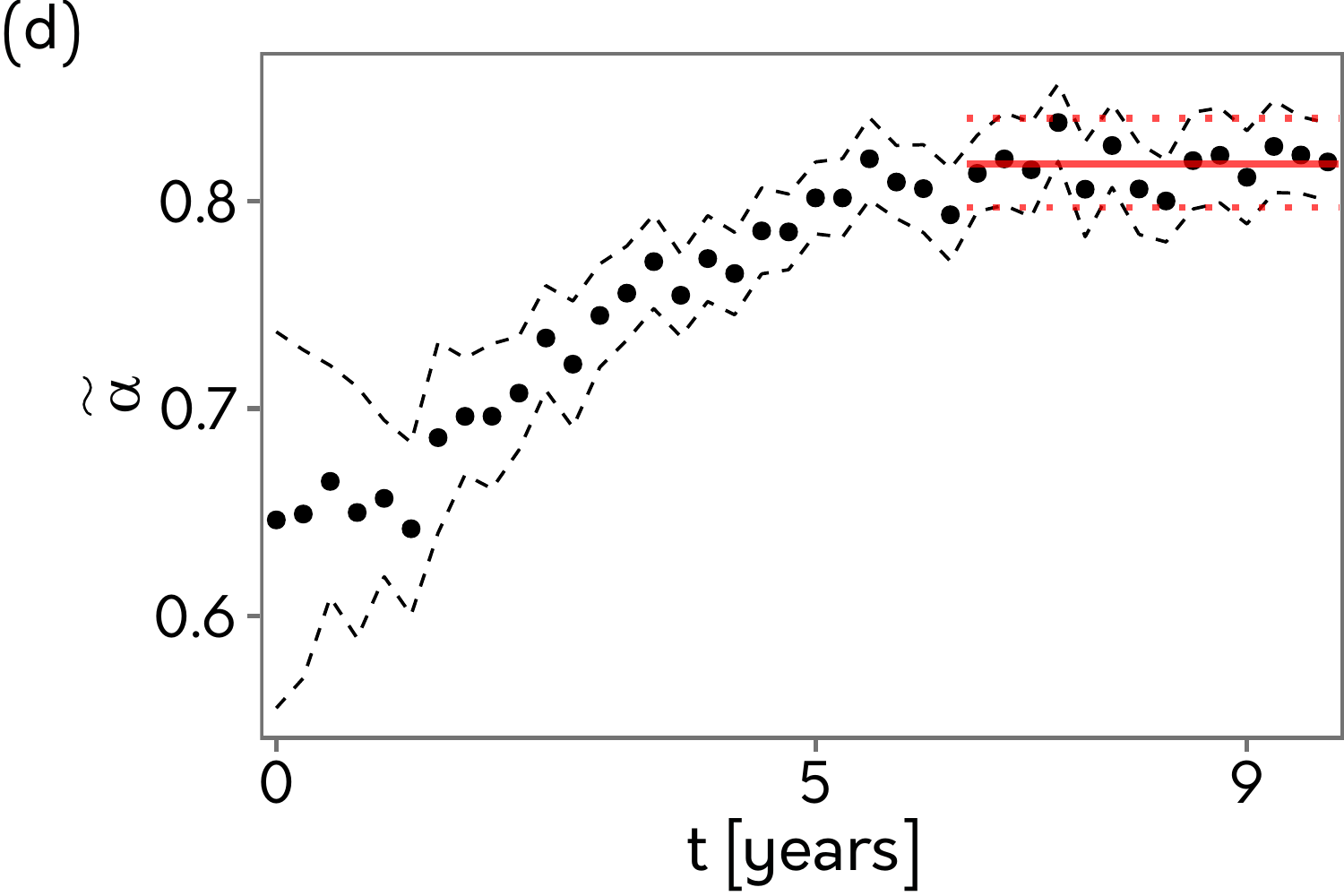}\hfill
\caption{\label{fig:fitIllu}%
Ageing and preferential attachment in the citation rate of USPTO patents
from CPC Section A granted during 1999-2001. Panel (a) shows fits of
exponential-aging functions to both the fitness-controlled citation rate
$\lambda_i/\eta_i\equiv [\Delta k_i(t+\Delta t)/\eta_i]/\Delta t$ (data
points indicated by triangles) and the uncontrolled-for-fitness average
citation rate $\bar\lambda_i \equiv \Delta k_i(t+\Delta t)/\Delta t$
(data points indicated by circles). Here $t$ corresponds to the age of
individual patents counted from the time of grant, and $\Delta t =
3\,$months. An example for typical fits of preferential-attachment
kernels to the data for both citation rates is provided in panel (b).
Panel (c) collates values for the obsolescence time $\tilde\tau$
extracted from fits of the r.h.s.\ of Eq.~(\ref{eq:fitConObs}) to
$\lambda_i/\eta_i$ for patents with fixed number $k$ of citations.
Similarly, panel (d) gives the exponents $\tilde\alpha$ obtained from
fitting the r.h.s.\ of Eq.~(\ref{eq:fitConPA}) to $\lambda_i/\eta_i$ for
patents with fixed age $t$. In panels (c) and (d), circles are fitted
parameter values, solid lines indicate their weighted averages, and the
black dashed (red dotted) curves show the 95\% confidence intervals for
fit-parameter values (their weighted averages).}
\end{figure*}

\textit{(ii)~Determining relative weighting.\/} The factors $\bar
u^{(b)}$ represent the minimally correlated variances of the
\textit{ad-hoc\/}-defined raw quality-indicator variables across a given
cohort of patents. They therefore approximate the truly independent
dimensions within which quality of individual patents can be measured by
observables available at the time of grant. To estimate the relative
importance of each factor in determining citation intensity, we perform
a forward-stepwise ordinary-least-squares regression on a small training
dataset with the Bayes Information Criterion~\cite{sch78} as our test
statistic and the log of the (mean-scaled) citations at grant as the
dependent variable~\footnote{We follow the usual convention where the
time of citation is taken to be the application date of the citing
patent~\cite{hal02}, whereas the age of cited patents is counted from the
time of grant~\cite{meh10}. Therefore, patents have often already accrued
citations by the time they are granted.}. Patents in the training dataset
are randomly selected and comprise about 10\% of a cohort, i.e.,
8,000-10,000 patents. The regression coefficients yield weights $w_b$
measuring the variances in the training dataset associated with the
factors $\bar u^{(b)}$. Figure~\ref{fig:EFAres}(b) shows the $w_b$
obtained for factors associated with the cohort of USPTO patents from CPC
Section A granted during 1999-2001~\cite{noteSM}.

\textit{(iii)~Achieving distributional fidelity.\/} By construction, the
linear combination of factors $\bar u^{(b)}$ with weights $w_b$ as
coefficients is a normally distributed quantity having zero mean and
variance given by $\sum_{b = 1}^{N_u} w_b^2$. Motivated by empirical
observations of citation rates~\cite{wan13}, we assume the distribution
$\rho(\eta)$ of fitnesses to be log-normal. For conceptual simplicity, we
fix the mean value of the fitness, $\mu_\eta \equiv 1$, which implies
$\mu_{\ln(\eta)} = -\sigma_{\ln(\eta)}^2/2$. Thus the only free parameter
characterizing the log-normal distribution of fitnesses is the standard
deviation $\sigma_{\ln(\eta)}$. It turns out to be possible to extract
$\sigma_{\ln(\eta)}$ from the observed dynamics of how patents acquire
their first citation without needing to assume anything explicit about
how the citation rate Eq.~(\ref{eq:fitRate}) depends on $k$ and $t$. More
specifically, we consider the time evolution of the fraction $n(0,t)$ of
uncited patents. Expanding its expression~\cite{hig17,gol18}
\begin{equation}
n(0, t) = \int d\eta\,\, \rho(\eta)\,\, \exp\left\{-\eta\, \tilde f(0)
\int_0^t d t^\prime\,\,\tilde A(t^\prime)\right\}
\end{equation}
in the short-time limit where $\tilde A(t)\approx \tilde A(0)$, and
assuming $\rho(\eta)$ to be log-normal, yields
\begin{equation}\label{eq:lnFitVar}
\ln n(0, t) \approx -\tilde A(0) \tilde f(0)\, t +
\frac{\mathrm{e}^{\sigma_{\ln(\eta)}^2} -1}{2} \left[ \tilde A(0)
\tilde f(0)\, t \right]^2 .
\end{equation}
Fitting Eq.~(\ref{eq:lnFitVar}) to the data enables us to extract
$\sigma_{\ln(\eta)}$ for each of the patent cohorts considered
here~\cite{noteSM}.

\begin{figure*}[t]
\includegraphics[width=0.49\textwidth]{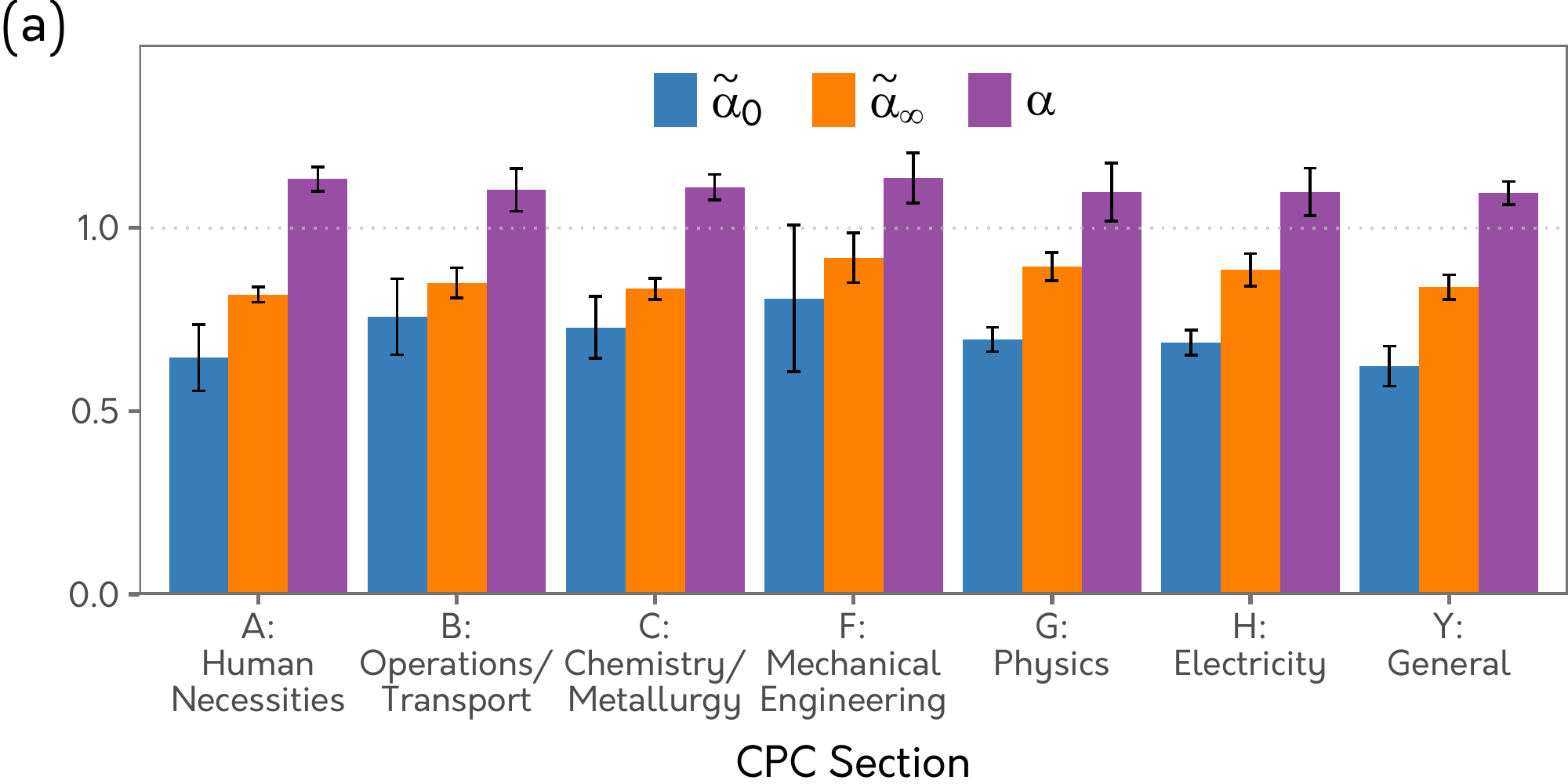}\hfill
\includegraphics[width=0.49\textwidth]{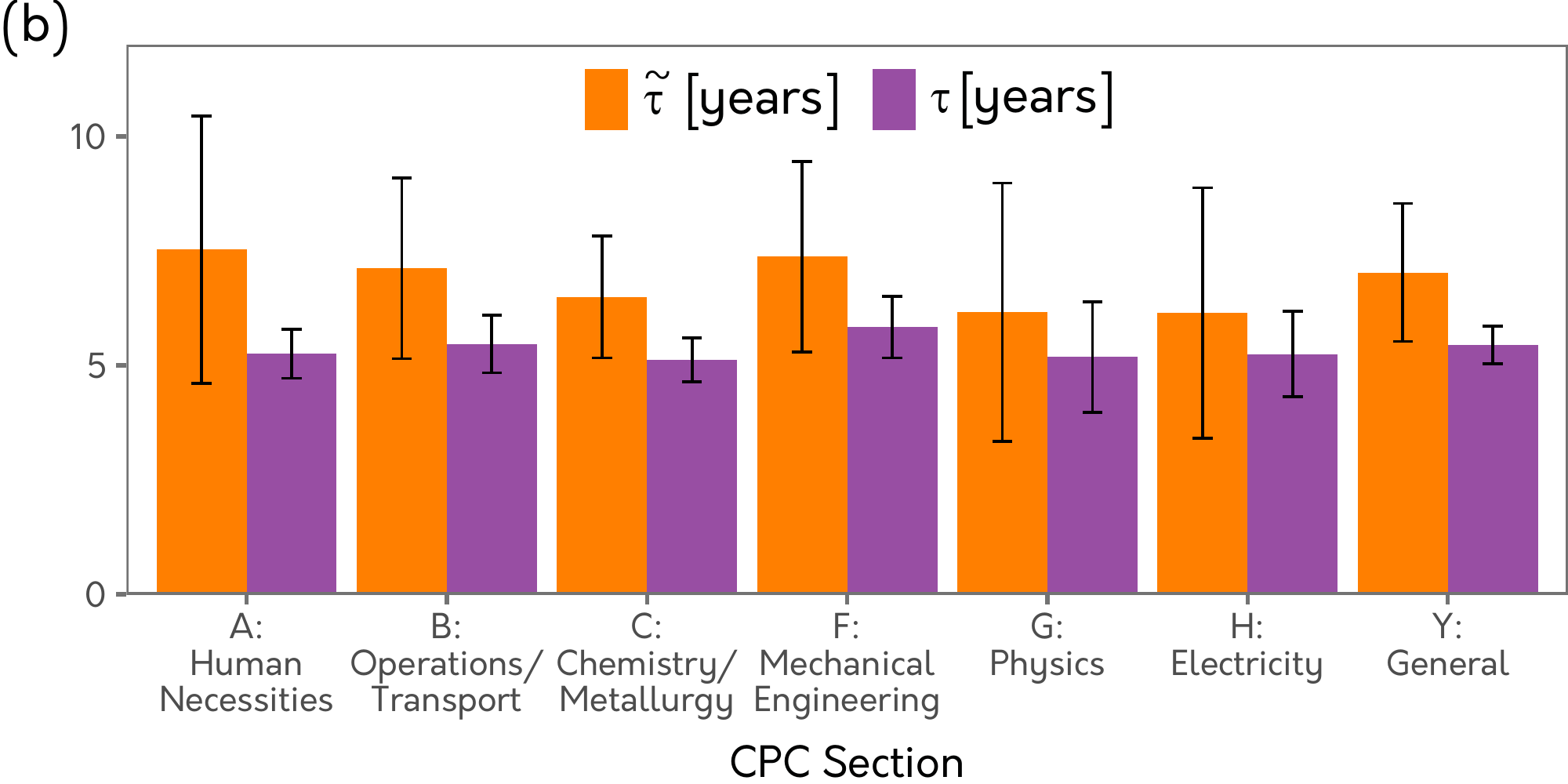}
\caption{\label{fig:results}%
Comparing preferential attachment and ageing exhibited in the
fitness-controlled citation rate $\lambda_i/\eta_i$ and the
uncontrolled-for-fitness average citation rate $\bar\lambda_i$,
respectively. Panel~(a) shows both the saturation value $\tilde
\alpha_\infty$ [the weighted average of the latest third among the values
obtained for $\tilde\alpha$ from fits of Eq.~(\ref{eq:fitConPA}) to
$\lambda_i/\eta_i$ for patent cohorts with fixed age $t$; see
Fig.~\ref{fig:fitIllu}(d)] and the value $\tilde\alpha_0$ for $t=0$
(corresponding to the time of grant) alongside $\alpha$ extracted from
fitting the uncontrolled-for-fitness average citation rate $\bar\lambda_i
\propto (k_i^\alpha + f_0)$. In panel~(b), the time $\tilde\tau$ ($\tau$)
is the weighted average of obsolescence times extracted from fitting
Eq.~(\ref{eq:fitConObs}) to fixed-$k$ bins of $\lambda_i/\eta_i$ as shown
in Fig.~\ref{fig:fitIllu}(c) for Section-A patents (fits of $\bar
\lambda_i \propto A_0\exp\{-t/\tau\}$). Citation data have been analyzed
for USPTO patents granted during 1999-2001 in the CPC Sections indicated
in the figure. Fitness values $\eta_i$ were estimated based on the
quality-indicator variables from Table~\ref{tab:qvars} using
Eq.~(\ref{eq:fitDef}) with Eq.~(\ref{eq:logVars}).}
\end{figure*}

Having addressed the three issues (i)--(iii) discussed above, we estimate
the intrinsic fitness of patent $i$ to attract citations in terms of the
observable values $v^{(a)}_i$ of the raw quality-indicator variables from
Table~\ref{tab:qvars} via
\begin{equation}\label{eq:fitDef}
\eta_i = \exp\left(-\frac{\sigma_{\ln(\eta)}^2}{2} + \frac{\sigma_{\ln
(\eta)}}{\sqrt{\sum_{b = 1}^{N_u} w_b^2}}\, \sum_{b=1}^{N_u} w_b\,
\sum_{a=1}^{N_v} L^{(b)}_a \, \bar v^{(a)}_i \right) \, .
\end{equation}
Assuming the rate at which citations are accrued by patents to follow the
expression (\ref{eq:fitRate}), we analyze the empirically observed
citation rates for USPTO patents from individual CPC Sections A, B, C, F,
G, H, and Y granted during 1999-2001. We allow for a ten-year time window
for citation accrual, starting from each individual patent's time of
grant. To account for citation inflation due to structural causes such as
fluctuations in R\&D spending or other policy decisions~\cite{kor99}, the
value of an incoming citation at time $t$ is scaled by the number of
patents applied for at that time~\cite{hig17,hig17b}. The change $\Delta
k_i(t + \Delta t)$ of inflation-adjusted citations received by patent $i$
over the time interval $(t, t+\Delta t]$ is then divided by the fitness
$\eta_i$ estimated for that patent in terms of its attributes $v^{(a)}_i$
according to Eq.~(\ref{eq:fitDef}). We find that the data for $[\Delta
k_i(t+\Delta t)/\eta_i]/\Delta t$ (corresponding to the
fitness-controlled citation rate $\lambda_i/\eta_i$) are best fitted in
terms of the product of an ageing function $\tilde A(t)$ and a
preferential-attachment kernel $\tilde f(k)$ given by
\begin{subequations}
\begin{eqnarray}\label{eq:fitConObs}
\tilde A(t) &=& \tilde A_0\,\exp\left( -\frac{t}{\tilde\tau} \right)
\quad , \\
\tilde f(k) &=& k^{\tilde\alpha} + \tilde f_0 \quad . \label{eq:fitConPA}
\end{eqnarray}
\end{subequations}
Figures~\ref{fig:fitIllu}(a) and \ref{fig:fitIllu}(b) show results
obtained for Section-A patents. For comparison, we include in the same
figures also fits of the quantity $\Delta k_i(t+\Delta t)/\Delta t$
representing the uncontrolled-for-fitness average citation rate $\bar
\lambda_i \equiv A(t)\, f(k_i)$ considered before~\cite{csa07,val07,
hig17}. In agreement with previous results~\cite{hig17} obtained using
the technology-classification scheme from Ref.~\cite{hal02}, we observe
$A(t)$ to be exponential in the long run while significantly exceeding
the exponential behavior extrapolated to short times. Fits to the data
for $\bar\lambda$ also establish the form~\cite{hig17} $f(k)=k^\alpha +
f_0$, with $\alpha > \tilde \alpha$.

Extracting the parameter $\tilde\tau$ corresponding to the time scale for
obsolescence from fits of the functional form given in
Eq.~(\ref{eq:fitConObs}) to the fitness-controlled citation rate $[\Delta
k_i(t+\Delta t)/\eta_i]/\Delta t$ for patents having a fixed number $k$
of citations yields results as shown for Section-A patents in
Fig.~\ref{fig:fitIllu}(c). The $\tilde\tau$ values fluctuate around a
mean value that is larger than that extracted from similar fits to the
uncontrolled-for-fitness citation rate. In contrast, as seen in
Fig.~\ref{fig:fitIllu}(d), the exponent $\tilde\alpha$ in the
preferential-attachment kernel Eq.~(\ref{eq:fitConPA}) obtained from
fitting this expression to the fitness-controlled citation rate for
patents at fixed age $t$ (counted from the time of grant) shows an
increasing trend at short times but eventually saturates.
Figure~\ref{fig:results} illustrates the obsolescence times and
preferential-attachment exponents extracted from citation data for the
patent cohorts considered in this work.

Comparing the dynamics of ageing and preferential attachment exhibited
by the fitness-controlled and uncontrolled-for-fitness citation rates,
respectively, reveals a number of interesting features. See
Fig.~\ref{fig:fitIllu}(a) for an illustration. Firstly, the exponential
time dependence from Eq.~(\ref{eq:fitConObs}) generally describes ageing
for the fitness-controlled citation rate very well at all times. In
contrast, the uncontrolled-for-fitness average citation rate at short
times systematically shows a large excess over the extrapolated
exponential behavior observed in the long-time limit~\cite{hig17,can19}.
Controlling for citation inflation as well as fitness has thus enabled us
to reveal more clearly the purely obsolescence-induced ageing process
governing patent-citation dynamics. (Deviations from the described
typical ageing behavior are observed sporadically~\cite{noteSM}.)
Generally, $\tilde\tau$ turns out to be larger by $\sim\! 1$--$2$ years
than the time scale $\tau$ extracted from the exponential ageing
displayed by the uncontrolled-for-fitness citation rate in the long-time
limit (see Fig.~\ref{fig:results}).

Another striking feature of the fitness-controlled citation rate is the
significantly reduced exponent characterizing preferential attachment,
especially at early stages of the citation process but persisting also
in the long-time limit. Theoretical studies~\cite{pha16,gol18} have
suggested that purely fitness-driven growth can cause phenomenologically
observed preferential-attachment dynamics. As our estimate of fitness
$\eta$ pertains to the attributes of patents known at the time of grant,
it can be expected that the mechanism for attracting citations at that
time is largely reflective of this fitness. We indeed find the exponent
$\tilde\alpha_0$ extracted from citations immediately after time of grant
to be much smaller than when fitness is not controlled for, suggesting
that a portion of the observed preferential attachment is due to
heterogeneous fitness. Even in the long-time limit, the fitness estimated
via quality indicators available at the time of grant still accounts for
a sizeable reduction by about 20-30\% in the exponent $\alpha$ governing
preferential attachment in the uncontrolled-for-fitness average citation
rate. More explicitly, the average fitness for patents having $k$
citations at time $t$ after grant is given by $\bar\eta(k, t) \equiv \bar
\lambda(k, t)/[\tilde A(t)\tilde f(k)]$ and constitutes a direct measure
for the fitness-explained fraction of the average citation rate
$\bar\lambda$. Our observations imply
\begin{equation}\label{eq:avFit}
\bar\eta(k, t) \to \frac{k^\alpha + f_0}{k^{\tilde\alpha_\infty} + \tilde
f_0} \, \frac{A_0}{\tilde A_0}\,\exp\left\{\frac{\tau-\tilde\tau}{\tau
\tilde\tau}\, t\right\}
\end{equation}
in the long-time limit. Thus $\bar\eta(k, t) \propto k^{\alpha-\tilde
\alpha_\infty}$ for patents having a citation count exceeding the average
in the long-time limit. The fact that $\alpha > \tilde\alpha_\infty$
implies that the most highly cited patents will be associated with, and
therefore detectable by, high values for the fitness variable. However,
although it turns out to be useful and predictive, $\eta_i$ as determined
by our procedure is still only a conservative and noisy estimate of the
true intrinsic fitness for patents to be cited. Identifying yet other, or
better, raw patent-quality indicators to include in the construction of
fitness could be a way to improve it further.

As more highly cited patents typically have greater fitness [cf.\
Eq.~(\ref{eq:avFit})], a weak trend of positive correlation between
obsolescence time and fitness can be deduced from
Fig.~\ref{fig:fitIllu}(c). This suggest that the commonly adopted
approach~\cite{bia01,gol18,pha16} where fitness enters the citation rate
(\ref{eq:fitRate}) as a prefactor likely constitutes only a first
approximation to a more complex interplay between fitness and ageing.
While it may be expected that better-quality patents age more slowly,
the level of noise in our data prevents firm conclusions to be drawn in
the present case.

In summary, we present an empirical study of the relation between basic
attributes of patents and their citation dynamics. Using only information
available at the time patents are granted, we are able to estimate their
intrinsic fitness for being cited. Even over time periods as long as 10
years afterwards, this fitness is found to determine significantly how
many more citations are accrued, especially for the most successful,
i.e., highly cited, inventions. Future research could extend our
construction of the fitness parameter governing citation dynamics to also
include attributes of the citing patents~\cite{fer12} and account
for social proximity between inventor communities~\cite{sor06}.


\begin{thebibliography}{51}%
\makeatletter
\providecommand \@ifxundefined [1]{%
 \@ifx{#1\undefined}
}%
\providecommand \@ifnum [1]{%
 \ifnum #1\expandafter \@firstoftwo
 \else \expandafter \@secondoftwo
 \fi
}%
\providecommand \@ifx [1]{%
 \ifx #1\expandafter \@firstoftwo
 \else \expandafter \@secondoftwo
 \fi
}%
\providecommand \natexlab [1]{#1}%
\providecommand \enquote  [1]{``#1''}%
\providecommand \bibnamefont  [1]{#1}%
\providecommand \bibfnamefont [1]{#1}%
\providecommand \citenamefont [1]{#1}%
\providecommand \href@noop [0]{\@secondoftwo}%
\providecommand \href [0]{\begingroup \@sanitize@url \@href}%
\providecommand \@href[1]{\@@startlink{#1}\@@href}%
\providecommand \@@href[1]{\endgroup#1\@@endlink}%
\providecommand \@sanitize@url [0]{\catcode `\\12\catcode `\$12\catcode
  `\&12\catcode `\#12\catcode `\^12\catcode `\_12\catcode `\%12\relax}%
\providecommand \@@startlink[1]{}%
\providecommand \@@endlink[0]{}%
\providecommand \url  [0]{\begingroup\@sanitize@url \@url }%
\providecommand \@url [1]{\endgroup\@href {#1}{\urlprefix }}%
\providecommand \urlprefix  [0]{URL }%
\providecommand \Eprint [0]{\href }%
\providecommand \doibase [0]{http://dx.doi.org/}%
\providecommand \selectlanguage [0]{\@gobble}%
\providecommand \bibinfo  [0]{\@secondoftwo}%
\providecommand \bibfield  [0]{\@secondoftwo}%
\providecommand \translation [1]{[#1]}%
\providecommand \BibitemOpen [0]{}%
\providecommand \bibitemStop [0]{}%
\providecommand \bibitemNoStop [0]{.\EOS\space}%
\providecommand \EOS [0]{\spacefactor3000\relax}%
\providecommand \BibitemShut  [1]{\csname bibitem#1\endcsname}%
\let\auto@bib@innerbib\@empty
\bibitem [{\citenamefont {D{\textquoteright}Souza}\ \emph
  {et~al.}(2007)\citenamefont {D{\textquoteright}Souza}, \citenamefont {Borgs},
  \citenamefont {Chayes}, \citenamefont {Berger},\ and\ \citenamefont
  {Kleinberg}}]{sou07}%
  \BibitemOpen
  \bibfield  {author} {\bibinfo {author} {\bibfnamefont {R.~M.}\ \bibnamefont
  {D{\textquoteright}Souza}}, \bibinfo {author} {\bibfnamefont
  {C.}~\bibnamefont {Borgs}}, \bibinfo {author} {\bibfnamefont {J.~T.}\
  \bibnamefont {Chayes}}, \bibinfo {author} {\bibfnamefont {N.}~\bibnamefont
  {Berger}}, \ and\ \bibinfo {author} {\bibfnamefont {R.~D.}\ \bibnamefont
  {Kleinberg}},\ }\href {\doibase 10.1073/pnas.0606779104} {\bibfield
  {journal} {\bibinfo  {journal} {Proc. Natl. Acad. Sci. U.S.A.}\ }\textbf
  {\bibinfo {volume} {104}},\ \bibinfo {pages} {6112} (\bibinfo {year}
  {2007})}\BibitemShut {NoStop}%
\bibitem [{\citenamefont {Kong}\ \emph {et~al.}(2008)\citenamefont {Kong},
  \citenamefont {Sarshar},\ and\ \citenamefont {Roychowdhury}}]{kon08}%
  \BibitemOpen
  \bibfield  {author} {\bibinfo {author} {\bibfnamefont {J.~S.}\ \bibnamefont
  {Kong}}, \bibinfo {author} {\bibfnamefont {N.}~\bibnamefont {Sarshar}}, \
  and\ \bibinfo {author} {\bibfnamefont {V.~P.}\ \bibnamefont {Roychowdhury}},\
  }\href {\doibase 10.1073/pnas.0805921105} {\bibfield  {journal} {\bibinfo
  {journal} {Proc. Natl. Acad. Sci. U.S.A.}\ }\textbf {\bibinfo {volume}
  {105}},\ \bibinfo {pages} {13724} (\bibinfo {year} {2008})}\BibitemShut
  {NoStop}%
\bibitem [{\citenamefont {Papadopoulos}\ \emph {et~al.}(2012)\citenamefont
  {Papadopoulos}, \citenamefont {Kitsak}, \citenamefont {Serrano},
  \citenamefont {Bogu{\~n}{\'a}},\ and\ \citenamefont {Krioukov}}]{pap12}%
  \BibitemOpen
  \bibfield  {author} {\bibinfo {author} {\bibfnamefont {F.}~\bibnamefont
  {Papadopoulos}}, \bibinfo {author} {\bibfnamefont {M.}~\bibnamefont
  {Kitsak}}, \bibinfo {author} {\bibfnamefont {M.~{\'A}.}\ \bibnamefont
  {Serrano}}, \bibinfo {author} {\bibfnamefont {M.}~\bibnamefont
  {Bogu{\~n}{\'a}}}, \ and\ \bibinfo {author} {\bibfnamefont {D.}~\bibnamefont
  {Krioukov}},\ }\href {\doibase 10.1038/nature11459} {\bibfield  {journal}
  {\bibinfo  {journal} {Nature}\ }\textbf {\bibinfo {volume} {489}},\ \bibinfo
  {pages} {537} (\bibinfo {year} {2012})}\BibitemShut {NoStop}%
\bibitem [{\citenamefont {Wang}\ \emph {et~al.}(2013)\citenamefont {Wang},
  \citenamefont {Song},\ and\ \citenamefont {Barab{\'a}si}}]{wan13}%
  \BibitemOpen
  \bibfield  {author} {\bibinfo {author} {\bibfnamefont {D.}~\bibnamefont
  {Wang}}, \bibinfo {author} {\bibfnamefont {C.}~\bibnamefont {Song}}, \ and\
  \bibinfo {author} {\bibfnamefont {A.-L.}\ \bibnamefont {Barab{\'a}si}},\
  }\href {\doibase 10.1126/science.1237825} {\bibfield  {journal} {\bibinfo
  {journal} {Science}\ }\textbf {\bibinfo {volume} {342}},\ \bibinfo {pages}
  {127} (\bibinfo {year} {2013})}\BibitemShut {NoStop}%
\bibitem [{\citenamefont {van~de Rijt}\ \emph {et~al.}(2014)\citenamefont
  {van~de Rijt}, \citenamefont {Kang}, \citenamefont {Restivo},\ and\
  \citenamefont {Patil}}]{rij14}%
  \BibitemOpen
  \bibfield  {author} {\bibinfo {author} {\bibfnamefont {A.}~\bibnamefont
  {van~de Rijt}}, \bibinfo {author} {\bibfnamefont {S.~M.}\ \bibnamefont
  {Kang}}, \bibinfo {author} {\bibfnamefont {M.}~\bibnamefont {Restivo}}, \
  and\ \bibinfo {author} {\bibfnamefont {A.}~\bibnamefont {Patil}},\ }\href
  {\doibase 10.1073/pnas.1316836111} {\bibfield  {journal} {\bibinfo  {journal}
  {Proc. Natl. Acad. Sci. U.S.A.}\ }\textbf {\bibinfo {volume} {111}},\
  \bibinfo {pages} {6934} (\bibinfo {year} {2014})}\BibitemShut {NoStop}%
\bibitem [{\citenamefont {Sinatra}\ \emph {et~al.}(2016)\citenamefont
  {Sinatra}, \citenamefont {Wang}, \citenamefont {Deville}, \citenamefont
  {Song},\ and\ \citenamefont {Barab{\'a}si}}]{sin16}%
  \BibitemOpen
  \bibfield  {author} {\bibinfo {author} {\bibfnamefont {R.}~\bibnamefont
  {Sinatra}}, \bibinfo {author} {\bibfnamefont {D.}~\bibnamefont {Wang}},
  \bibinfo {author} {\bibfnamefont {P.}~\bibnamefont {Deville}}, \bibinfo
  {author} {\bibfnamefont {C.}~\bibnamefont {Song}}, \ and\ \bibinfo {author}
  {\bibfnamefont {A.-L.}\ \bibnamefont {Barab{\'a}si}},\ }\href {\doibase
  10.1126/science.aaf5239} {\bibfield  {journal} {\bibinfo  {journal}
  {Science}\ }\textbf {\bibinfo {volume} {354}},\ \bibinfo {pages} {aaf5239}
  (\bibinfo {year} {2016})}\BibitemShut {NoStop}%
\bibitem [{\citenamefont {Zeng}\ \emph {et~al.}(2017)\citenamefont {Zeng},
  \citenamefont {Shen}, \citenamefont {Zhou}, \citenamefont {Wu}, \citenamefont
  {Fan}, \citenamefont {Wang},\ and\ \citenamefont {Stanley}}]{zen17}%
  \BibitemOpen
  \bibfield  {author} {\bibinfo {author} {\bibfnamefont {A.}~\bibnamefont
  {Zeng}}, \bibinfo {author} {\bibfnamefont {Z.}~\bibnamefont {Shen}}, \bibinfo
  {author} {\bibfnamefont {J.}~\bibnamefont {Zhou}}, \bibinfo {author}
  {\bibfnamefont {J.}~\bibnamefont {Wu}}, \bibinfo {author} {\bibfnamefont
  {Y.}~\bibnamefont {Fan}}, \bibinfo {author} {\bibfnamefont {Y.}~\bibnamefont
  {Wang}}, \ and\ \bibinfo {author} {\bibfnamefont {H.~E.}\ \bibnamefont
  {Stanley}},\ }\href {\doibase 10.1016/j.physrep.2017.10.001} {\bibfield
  {journal} {\bibinfo  {journal} {Phys. Repts.}\ }\textbf {\bibinfo {volume}
  {714-715}},\ \bibinfo {pages} {1} (\bibinfo {year} {2017})}\BibitemShut
  {NoStop}%
\bibitem [{\citenamefont {Fortunato}\ \emph {et~al.}(2018)\citenamefont
  {Fortunato}, \citenamefont {Bergstrom}, \citenamefont {B{\"o}rner},
  \citenamefont {Evans}, \citenamefont {Helbing}, \citenamefont
  {Milojevi{\'c}}, \citenamefont {Petersen}, \citenamefont {Radicchi},
  \citenamefont {Sinatra}, \citenamefont {Uzzi}, \citenamefont {Vespignani},
  \citenamefont {Waltman}, \citenamefont {Wang},\ and\ \citenamefont
  {Barab{\'a}si}}]{for18}%
  \BibitemOpen
  \bibfield  {author} {\bibinfo {author} {\bibfnamefont {S.}~\bibnamefont
  {Fortunato}}, \bibinfo {author} {\bibfnamefont {C.~T.}\ \bibnamefont
  {Bergstrom}}, \bibinfo {author} {\bibfnamefont {K.}~\bibnamefont
  {B{\"o}rner}}, \bibinfo {author} {\bibfnamefont {J.~A.}\ \bibnamefont
  {Evans}}, \bibinfo {author} {\bibfnamefont {D.}~\bibnamefont {Helbing}},
  \bibinfo {author} {\bibfnamefont {S.}~\bibnamefont {Milojevi{\'c}}}, \bibinfo
  {author} {\bibfnamefont {A.~M.}\ \bibnamefont {Petersen}}, \bibinfo {author}
  {\bibfnamefont {F.}~\bibnamefont {Radicchi}}, \bibinfo {author}
  {\bibfnamefont {R.}~\bibnamefont {Sinatra}}, \bibinfo {author} {\bibfnamefont
  {B.}~\bibnamefont {Uzzi}}, \bibinfo {author} {\bibfnamefont {A.}~\bibnamefont
  {Vespignani}}, \bibinfo {author} {\bibfnamefont {L.}~\bibnamefont {Waltman}},
  \bibinfo {author} {\bibfnamefont {D.}~\bibnamefont {Wang}}, \ and\ \bibinfo
  {author} {\bibfnamefont {A.-L.}\ \bibnamefont {Barab{\'a}si}},\ }\href
  {\doibase 10.1126/science.aao0185} {\bibfield  {journal} {\bibinfo  {journal}
  {Science}\ }\textbf {\bibinfo {volume} {359}},\ \bibinfo {pages} {eaao0185}
  (\bibinfo {year} {2018})}\BibitemShut {NoStop}%
\bibitem [{\citenamefont {Mariani}\ \emph {et~al.}(2018)\citenamefont
  {Mariani}, \citenamefont {Medo},\ and\ \citenamefont {Lafond}}]{mar18}%
  \BibitemOpen
  \bibfield  {author} {\bibinfo {author} {\bibfnamefont {M.~S.}\ \bibnamefont
  {Mariani}}, \bibinfo {author} {\bibfnamefont {M.}~\bibnamefont {Medo}}, \
  and\ \bibinfo {author} {\bibfnamefont {F.}~\bibnamefont {Lafond}},\ }\href
  {\doibase 10.1016/j.techfore.2018.01.036} {\bibfield  {journal} {\bibinfo
  {journal} {Technol. Forecast. Soc. Change}\ }\textbf {\bibinfo {volume}
  {\hspace{0pt}}},\ \bibinfo {pages} {\hspace{0pt}} (\bibinfo {year} {2018})},\
  \bibinfo {note} {in press}\BibitemShut {NoStop}%
\bibitem [{\citenamefont {Vespignani}(2012)}]{ves12}%
  \BibitemOpen
  \bibfield  {author} {\bibinfo {author} {\bibfnamefont {A.}~\bibnamefont
  {Vespignani}},\ }\href {\doibase 10.1038/nphys2160} {\bibfield  {journal}
  {\bibinfo  {journal} {Nat. Phys.}\ }\textbf {\bibinfo {volume} {8}},\
  \bibinfo {pages} {32} (\bibinfo {year} {2012})}\BibitemShut {NoStop}%
\bibitem [{\citenamefont {Jaffe}\ and\ \citenamefont {{de
  Rassenfosse}}(2017)}]{jaf17}%
  \BibitemOpen
  \bibfield  {author} {\bibinfo {author} {\bibfnamefont {A.~B.}\ \bibnamefont
  {Jaffe}}\ and\ \bibinfo {author} {\bibfnamefont {G.}~\bibnamefont {{de
  Rassenfosse}}},\ }\href {\doibase 10.1002/asi.23731} {\bibfield  {journal}
  {\bibinfo  {journal} {J. Assoc. Inf. Sci. Technol.}\ }\textbf {\bibinfo
  {volume} {68}},\ \bibinfo {pages} {1360} (\bibinfo {year}
  {2017})}\BibitemShut {NoStop}%
\bibitem [{\citenamefont {Higham}\ \emph
  {et~al.}(2017{\natexlab{a}})\citenamefont {Higham}, \citenamefont
  {Governale}, \citenamefont {Jaffe},\ and\ \citenamefont {Z\"ulicke}}]{hig17}%
  \BibitemOpen
  \bibfield  {author} {\bibinfo {author} {\bibfnamefont {K.~W.}\ \bibnamefont
  {Higham}}, \bibinfo {author} {\bibfnamefont {M.}~\bibnamefont {Governale}},
  \bibinfo {author} {\bibfnamefont {A.~B.}\ \bibnamefont {Jaffe}}, \ and\
  \bibinfo {author} {\bibfnamefont {U.}~\bibnamefont {Z\"ulicke}},\ }\href
  {\doibase 10.1103/PhysRevE.95.042309} {\bibfield  {journal} {\bibinfo
  {journal} {Phys. Rev. E}\ }\textbf {\bibinfo {volume} {95}},\ \bibinfo
  {pages} {042309} (\bibinfo {year} {2017}{\natexlab{a}})}\BibitemShut
  {NoStop}%
\bibitem [{\citenamefont {Hall}\ \emph {et~al.}(2002)\citenamefont {Hall},
  \citenamefont {Jaffe},\ and\ \citenamefont {Trajtenberg}}]{hal02}%
  \BibitemOpen
  \bibfield  {author} {\bibinfo {author} {\bibfnamefont {B.~H.}\ \bibnamefont
  {Hall}}, \bibinfo {author} {\bibfnamefont {A.~B.}\ \bibnamefont {Jaffe}}, \
  and\ \bibinfo {author} {\bibfnamefont {M.}~\bibnamefont {Trajtenberg}},\ }in\
  \href@noop {} {\emph {\bibinfo {booktitle} {Patents, Citations, and
  Innovations: A Window on the Knowledge Economy}}},\ \bibinfo {editor} {edited
  by\ \bibinfo {editor} {\bibfnamefont {A.~B.}\ \bibnamefont {Jaffe}}\ and\
  \bibinfo {editor} {\bibfnamefont {M.}~\bibnamefont {Trajtenberg}}}\ (\bibinfo
   {publisher} {MIT Press},\ \bibinfo {address} {Cambridge, MA},\ \bibinfo
  {year} {2002})\ pp.\ \bibinfo {pages} {403--460}\BibitemShut {NoStop}%
\bibitem [{\citenamefont {Cs{\'a}rdi}\ \emph {et~al.}(2007)\citenamefont
  {Cs{\'a}rdi}, \citenamefont {Strandburg}, \citenamefont {Zal{\'a}nyi},
  \citenamefont {Tobochnik},\ and\ \citenamefont {{\'E}rdi}}]{csa07}%
  \BibitemOpen
  \bibfield  {author} {\bibinfo {author} {\bibfnamefont {G.}~\bibnamefont
  {Cs{\'a}rdi}}, \bibinfo {author} {\bibfnamefont {K.~J.}\ \bibnamefont
  {Strandburg}}, \bibinfo {author} {\bibfnamefont {L.}~\bibnamefont
  {Zal{\'a}nyi}}, \bibinfo {author} {\bibfnamefont {J.}~\bibnamefont
  {Tobochnik}}, \ and\ \bibinfo {author} {\bibfnamefont {P.}~\bibnamefont
  {{\'E}rdi}},\ }\href {\doibase 10.1016/j.physa.2006.08.022} {\bibfield
  {journal} {\bibinfo  {journal} {Physica A}\ }\textbf {\bibinfo {volume}
  {374}},\ \bibinfo {pages} {783} (\bibinfo {year} {2007})}\BibitemShut
  {NoStop}%
\bibitem [{\citenamefont {Valverde}\ \emph {et~al.}(2007)\citenamefont
  {Valverde}, \citenamefont {Sol{\'e}}, \citenamefont {Bedau},\ and\
  \citenamefont {Packard}}]{val07}%
  \BibitemOpen
  \bibfield  {author} {\bibinfo {author} {\bibfnamefont {S.}~\bibnamefont
  {Valverde}}, \bibinfo {author} {\bibfnamefont {R.~V.}\ \bibnamefont
  {Sol{\'e}}}, \bibinfo {author} {\bibfnamefont {M.~A.}\ \bibnamefont {Bedau}},
  \ and\ \bibinfo {author} {\bibfnamefont {N.}~\bibnamefont {Packard}},\ }\href
  {\doibase 10.1103/PhysRevE.76.056118} {\bibfield  {journal} {\bibinfo
  {journal} {Phys. Rev. E}\ }\textbf {\bibinfo {volume} {76}},\ \bibinfo
  {pages} {056118} (\bibinfo {year} {2007})}\BibitemShut {NoStop}%
\bibitem [{\citenamefont {Dorogovtsev}\ and\ \citenamefont
  {Mendes}(2000)}]{dor00}%
  \BibitemOpen
  \bibfield  {author} {\bibinfo {author} {\bibfnamefont {S.~N.}\ \bibnamefont
  {Dorogovtsev}}\ and\ \bibinfo {author} {\bibfnamefont {J.~F.~F.}\
  \bibnamefont {Mendes}},\ }\href {\doibase 10.1103/PhysRevE.62.1842}
  {\bibfield  {journal} {\bibinfo  {journal} {Phys. Rev. E}\ }\textbf {\bibinfo
  {volume} {62}},\ \bibinfo {pages} {1842} (\bibinfo {year}
  {2000})}\BibitemShut {NoStop}%
\bibitem [{\citenamefont {Price}(1976)}]{pri76}%
  \BibitemOpen
  \bibfield  {author} {\bibinfo {author} {\bibfnamefont {D.~d.~S.}\
  \bibnamefont {Price}},\ }\href {\doibase 10.1002/asi.4630270505} {\bibfield
  {journal} {\bibinfo  {journal} {J. Am. Soc. Inf. Sci.}\ }\textbf {\bibinfo
  {volume} {27}},\ \bibinfo {pages} {292} (\bibinfo {year} {1976})}\BibitemShut
  {NoStop}%
\bibitem [{\citenamefont {Barab{\'a}si}\ and\ \citenamefont
  {Albert}(1999)}]{bar99}%
  \BibitemOpen
  \bibfield  {author} {\bibinfo {author} {\bibfnamefont {A.-L.}\ \bibnamefont
  {Barab{\'a}si}}\ and\ \bibinfo {author} {\bibfnamefont {R.}~\bibnamefont
  {Albert}},\ }\href {\doibase 10.1126/science.286.5439.509} {\bibfield
  {journal} {\bibinfo  {journal} {Science}\ }\textbf {\bibinfo {volume}
  {286}},\ \bibinfo {pages} {509} (\bibinfo {year} {1999})}\BibitemShut
  {NoStop}%
\bibitem [{\citenamefont {Krapivsky}\ and\ \citenamefont
  {Redner}(2001)}]{kra01}%
  \BibitemOpen
  \bibfield  {author} {\bibinfo {author} {\bibfnamefont {P.~L.}\ \bibnamefont
  {Krapivsky}}\ and\ \bibinfo {author} {\bibfnamefont {S.}~\bibnamefont
  {Redner}},\ }\href {\doibase 10.1103/PhysRevE.63.066123} {\bibfield
  {journal} {\bibinfo  {journal} {Phys. Rev. E}\ }\textbf {\bibinfo {volume}
  {63}},\ \bibinfo {pages} {066123} (\bibinfo {year} {2001})}\BibitemShut
  {NoStop}%
\bibitem [{\citenamefont {Redner}(2005)}]{red05}%
  \BibitemOpen
  \bibfield  {author} {\bibinfo {author} {\bibfnamefont {S.}~\bibnamefont
  {Redner}},\ }\href {\doibase 10.1063/1.1996475} {\bibfield  {journal}
  {\bibinfo  {journal} {Phys. Today}\ }\textbf {\bibinfo {volume} {58}},\
  \bibinfo {pages} {49} (\bibinfo {year} {2005})}\BibitemShut {NoStop}%
\bibitem [{\citenamefont {Golosovsky}\ and\ \citenamefont
  {Solomon}(2012)}]{gol12}%
  \BibitemOpen
  \bibfield  {author} {\bibinfo {author} {\bibfnamefont {M.}~\bibnamefont
  {Golosovsky}}\ and\ \bibinfo {author} {\bibfnamefont {S.}~\bibnamefont
  {Solomon}},\ }\href {\doibase 10.1103/PhysRevLett.109.098701} {\bibfield
  {journal} {\bibinfo  {journal} {Phys. Rev. Lett.}\ }\textbf {\bibinfo
  {volume} {109}},\ \bibinfo {pages} {098701} (\bibinfo {year}
  {2012})}\BibitemShut {NoStop}%
\bibitem [{\citenamefont {Higham}\ \emph
  {et~al.}(2017{\natexlab{b}})\citenamefont {Higham}, \citenamefont
  {Governale}, \citenamefont {Jaffe},\ and\ \citenamefont
  {Z{\"u}licke}}]{hig17b}%
  \BibitemOpen
  \bibfield  {author} {\bibinfo {author} {\bibfnamefont {K.}~\bibnamefont
  {Higham}}, \bibinfo {author} {\bibfnamefont {M.}~\bibnamefont {Governale}},
  \bibinfo {author} {\bibfnamefont {A.}~\bibnamefont {Jaffe}}, \ and\ \bibinfo
  {author} {\bibfnamefont {U.}~\bibnamefont {Z{\"u}licke}},\ }\href {\doibase
  10.1016/j.joi.2017.10.004} {\bibfield  {journal} {\bibinfo  {journal} {J.
  Informetrics}\ }\textbf {\bibinfo {volume} {11}},\ \bibinfo {pages} {1190}
  (\bibinfo {year} {2017}{\natexlab{b}})}\BibitemShut {NoStop}%
\bibitem [{Note1()}]{Note1}%
  \BibitemOpen
  \bibinfo {note} {Qualitative~\cite {mey00} and quantitative~\cite {clo15}
  studies have shown that patent citations are less likely to be irrelevant or
  superfluous. Also, unlike scientific articles, patent publications contain a
  highly regulated set of metadata suitable for constructing universal
  intrinsic-quality indicators.}\BibitemShut {Stop}%
\bibitem [{\citenamefont {Marco}(2007)}]{mar07}%
  \BibitemOpen
  \bibfield  {author} {\bibinfo {author} {\bibfnamefont {A.~C.}\ \bibnamefont
  {Marco}},\ }\href {\doibase 10.1016/j.econlet.2006.08.014} {\bibfield
  {journal} {\bibinfo  {journal} {Econom. Lett.}\ }\textbf {\bibinfo {volume}
  {94}},\ \bibinfo {pages} {290} (\bibinfo {year} {2007})}\BibitemShut
  {NoStop}%
\bibitem [{\citenamefont {Bianconi}\ and\ \citenamefont
  {Barab{\'a}si}(2001)}]{bia01}%
  \BibitemOpen
  \bibfield  {author} {\bibinfo {author} {\bibfnamefont {G.}~\bibnamefont
  {Bianconi}}\ and\ \bibinfo {author} {\bibfnamefont {A.-L.}\ \bibnamefont
  {Barab{\'a}si}},\ }\href {\doibase 10.1209/epl/i2001-00260-6} {\bibfield
  {journal} {\bibinfo  {journal} {Europhys. Lett.}\ }\textbf {\bibinfo {volume}
  {54}},\ \bibinfo {pages} {436} (\bibinfo {year} {2001})}\BibitemShut
  {NoStop}%
\bibitem [{Note2()}]{Note2}%
  \BibitemOpen
  \bibinfo {note} {Without loss of generality, we assume fitness $\eta $ to be
  normalized such that its mean satisfies $\mu _\eta =1$.}\BibitemShut {Stop}%
\bibitem [{\citenamefont {Ferretti}\ \emph {et~al.}(2012)\citenamefont
  {Ferretti}, \citenamefont {Cortelezzi}, \citenamefont {Yang}, \citenamefont
  {Marmorini},\ and\ \citenamefont {Bianconi}}]{fer12}%
  \BibitemOpen
  \bibfield  {author} {\bibinfo {author} {\bibfnamefont {L.}~\bibnamefont
  {Ferretti}}, \bibinfo {author} {\bibfnamefont {M.}~\bibnamefont
  {Cortelezzi}}, \bibinfo {author} {\bibfnamefont {B.}~\bibnamefont {Yang}},
  \bibinfo {author} {\bibfnamefont {G.}~\bibnamefont {Marmorini}}, \ and\
  \bibinfo {author} {\bibfnamefont {G.}~\bibnamefont {Bianconi}},\ }\href
  {\doibase 10.1103/PhysRevE.85.066110} {\bibfield  {journal} {\bibinfo
  {journal} {Phys. Rev. E}\ }\textbf {\bibinfo {volume} {85}},\ \bibinfo
  {pages} {066110} (\bibinfo {year} {2012})}\BibitemShut {NoStop}%
\bibitem [{\citenamefont {Golosovsky}(2018)}]{gol18}%
  \BibitemOpen
  \bibfield  {author} {\bibinfo {author} {\bibfnamefont {M.}~\bibnamefont
  {Golosovsky}},\ }\href {\doibase 10.1103/PhysRevE.97.062310} {\bibfield
  {journal} {\bibinfo  {journal} {Phys. Rev. E}\ }\textbf {\bibinfo {volume}
  {97}},\ \bibinfo {pages} {062310} (\bibinfo {year} {2018})}\BibitemShut
  {NoStop}%
\bibitem [{\citenamefont {Pham}\ \emph {et~al.}(2016)\citenamefont {Pham},
  \citenamefont {Sheridan},\ and\ \citenamefont {Shimodaira}}]{pha16}%
  \BibitemOpen
  \bibfield  {author} {\bibinfo {author} {\bibfnamefont {T.}~\bibnamefont
  {Pham}}, \bibinfo {author} {\bibfnamefont {P.}~\bibnamefont {Sheridan}}, \
  and\ \bibinfo {author} {\bibfnamefont {H.}~\bibnamefont {Shimodaira}},\
  }\href {\doibase 10.1038/srep32558} {\bibfield  {journal} {\bibinfo
  {journal} {Sci. Rep.}\ }\textbf {\bibinfo {volume} {6}},\ \bibinfo {pages}
  {440} (\bibinfo {year} {2016})}\BibitemShut {NoStop}%
\bibitem [{\citenamefont {Newman}\ and\ \citenamefont {Leicht}(2007)}]{new07}%
  \BibitemOpen
  \bibfield  {author} {\bibinfo {author} {\bibfnamefont {M.~E.~J.}\
  \bibnamefont {Newman}}\ and\ \bibinfo {author} {\bibfnamefont {E.~A.}\
  \bibnamefont {Leicht}},\ }\href {\doibase 10.1073/pnas.0610537104} {\bibfield
   {journal} {\bibinfo  {journal} {Proc. Natl. Acad. Sci. U.S.A.}\ }\textbf
  {\bibinfo {volume} {104}},\ \bibinfo {pages} {9564} (\bibinfo {year}
  {2007})}\BibitemShut {NoStop}%
\bibitem [{\citenamefont {Ronda-Pupo}\ and\ \citenamefont
  {Pham}(2018)}]{ron18}%
  \BibitemOpen
  \bibfield  {author} {\bibinfo {author} {\bibfnamefont {G.~A.}\ \bibnamefont
  {Ronda-Pupo}}\ and\ \bibinfo {author} {\bibfnamefont {T.}~\bibnamefont
  {Pham}},\ }\href {\doibase 10.1007/s11192-018-2761-3} {\bibfield  {journal}
  {\bibinfo  {journal} {Scientometrics}\ }\textbf {\bibinfo {volume} {116}},\
  \bibinfo {pages} {363} (\bibinfo {year} {2018})}\BibitemShut {NoStop}%
\bibitem [{Note3()}]{Note3}%
  \BibitemOpen
  \bibinfo {note} {A rate of the form given in Eq.~(\ref {eq:fitRate}) covers
  the extreme cases of growth by pure preferential attachment~\cite {bar99} [by
  letting $\rho (\eta )\to \delta (\eta - 1)$, where $\delta (\cdot )$ is the
  Dirac-$\delta $ function] or pure fitness~\cite {cal02,gar04,tac12}
  [$\protect \mathaccentV {tilde}07Ef(k)\to \protect \mathrm
  {constant}$].}\BibitemShut {Stop}%
\bibitem [{\citenamefont {Lanjouw}\ and\ \citenamefont
  {Schankerman}(2004)}]{lan04}%
  \BibitemOpen
  \bibfield  {author} {\bibinfo {author} {\bibfnamefont {J.~O.}\ \bibnamefont
  {Lanjouw}}\ and\ \bibinfo {author} {\bibfnamefont {M.}~\bibnamefont
  {Schankerman}},\ }\href {\doibase 10.1111/j.1468-0297.2004.00216.x}
  {\bibfield  {journal} {\bibinfo  {journal} {Econom. J.}\ }\textbf {\bibinfo
  {volume} {114}},\ \bibinfo {pages} {441} (\bibinfo {year}
  {2004})}\BibitemShut {NoStop}%
\bibitem [{\citenamefont {Gambardella}\ \emph {et~al.}(2008)\citenamefont
  {Gambardella}, \citenamefont {Harhoff},\ and\ \citenamefont
  {Verspagen}}]{gam08}%
  \BibitemOpen
  \bibfield  {author} {\bibinfo {author} {\bibfnamefont {A.}~\bibnamefont
  {Gambardella}}, \bibinfo {author} {\bibfnamefont {D.}~\bibnamefont
  {Harhoff}}, \ and\ \bibinfo {author} {\bibfnamefont {B.}~\bibnamefont
  {Verspagen}},\ }\href {\doibase 10.1057/emr.2008.10} {\bibfield  {journal}
  {\bibinfo  {journal} {Eur. Manage. Rev.}\ }\textbf {\bibinfo {volume} {5}},\
  \bibinfo {pages} {69} (\bibinfo {year} {2008})}\BibitemShut {NoStop}%
\bibitem [{\citenamefont {{van Zeebroeck}}\ and\ \citenamefont {{van
  Pottelsberghe de la Potterie}}(2011)}]{zee11}%
  \BibitemOpen
  \bibfield  {author} {\bibinfo {author} {\bibfnamefont {N.}~\bibnamefont {{van
  Zeebroeck}}}\ and\ \bibinfo {author} {\bibfnamefont {B.}~\bibnamefont {{van
  Pottelsberghe de la Potterie}}},\ }\href {\doibase 10.1080/10438591003668638}
  {\bibfield  {journal} {\bibinfo  {journal} {Econ. Innov. New Technol.}\
  }\textbf {\bibinfo {volume} {20}},\ \bibinfo {pages} {283} (\bibinfo {year}
  {2011})}\BibitemShut {NoStop}%
\bibitem [{\citenamefont {Kogan}\ \emph {et~al.}(2017)\citenamefont {Kogan},
  \citenamefont {Papanikolaou}, \citenamefont {Seru},\ and\ \citenamefont
  {Stoffman}}]{kog17}%
  \BibitemOpen
  \bibfield  {author} {\bibinfo {author} {\bibfnamefont {L.}~\bibnamefont
  {Kogan}}, \bibinfo {author} {\bibfnamefont {D.}~\bibnamefont {Papanikolaou}},
  \bibinfo {author} {\bibfnamefont {A.}~\bibnamefont {Seru}}, \ and\ \bibinfo
  {author} {\bibfnamefont {N.}~\bibnamefont {Stoffman}},\ }\href {\doibase
  10.1093/qje/qjw040} {\bibfield  {journal} {\bibinfo  {journal} {Quart. J.
  Econom.}\ }\textbf {\bibinfo {volume} {132}},\ \bibinfo {pages} {665}
  (\bibinfo {year} {2017})}\BibitemShut {NoStop}%
\bibitem [{\citenamefont {{de Rassenfosse}}\ and\ \citenamefont
  {Jaffe}(2018)}]{ras18}%
  \BibitemOpen
  \bibfield  {author} {\bibinfo {author} {\bibfnamefont {G.}~\bibnamefont {{de
  Rassenfosse}}}\ and\ \bibinfo {author} {\bibfnamefont {A.~B.}\ \bibnamefont
  {Jaffe}},\ }\href {\doibase 10.1111/jems.12219} {\bibfield  {journal}
  {\bibinfo  {journal} {J. Econ. Manage. Strategy}\ }\textbf {\bibinfo {volume}
  {27}},\ \bibinfo {pages} {134} (\bibinfo {year} {2018})}\BibitemShut
  {NoStop}%
\bibitem [{\citenamefont {Mulaik}(2010)}]{mul10}%
  \BibitemOpen
  \bibfield  {author} {\bibinfo {author} {\bibfnamefont {S.~A.}\ \bibnamefont
  {Mulaik}},\ }\href@noop {} {\emph {\bibinfo {title} {Foundations of Factor
  Analysis}}},\ \bibinfo {edition} {2nd}\ ed.\ (\bibinfo  {publisher} {CRC
  Press},\ \bibinfo {address} {Boca Raton},\ \bibinfo {year}
  {2010})\BibitemShut {NoStop}%
\bibitem [{not()}]{noteSM}%
  \BibitemOpen
  \href@noop {} {}\bibinfo {note} {More specific details about patent-quality
  variables and statistical methods, as well as instructive additional results,
  are provided in the
  \href{http://doi.org/10.5281/zenodo.2564440}{Supplemental Material}
  (DOI: 10.5281/zenodo.2564440).}\BibitemShut {Stop}%
\bibitem [{Note4()}]{Note4}%
  \BibitemOpen
  \bibinfo {note} {We use $\mu _q$ and $\sigma _q$ to indicate the mean and
  variance, respectively, of any randomly distributed quantity
  $q$.}\BibitemShut {Stop}%
\bibitem [{\citenamefont {Schwarz}(1978)}]{sch78}%
  \BibitemOpen
  \bibfield  {author} {\bibinfo {author} {\bibfnamefont {G.}~\bibnamefont
  {Schwarz}},\ }\href {\doibase 10.1214/aos/1176344136} {\bibfield  {journal}
  {\bibinfo  {journal} {Ann. Statist.}\ }\textbf {\bibinfo {volume} {6}},\
  \bibinfo {pages} {461} (\bibinfo {year} {1978})}\BibitemShut {NoStop}%
\bibitem [{Note5()}]{Note5}%
  \BibitemOpen
  \bibinfo {note} {We follow the usual convention where the time of citation is
  taken to be the application date of the citing patent~\cite {hal02}, whereas
  the age of cited patents is counted from the time of grant~\cite {meh10}.
  Therefore, patents have often already accrued citations by the time they are
  granted.}\BibitemShut {Stop}%
\bibitem [{\citenamefont {Kortum}\ and\ \citenamefont {Lerner}(1999)}]{kor99}%
  \BibitemOpen
  \bibfield  {author} {\bibinfo {author} {\bibfnamefont {S.}~\bibnamefont
  {Kortum}}\ and\ \bibinfo {author} {\bibfnamefont {J.}~\bibnamefont
  {Lerner}},\ }\href {\doibase 10.1016/S0048-7333(98)00082-1} {\bibfield
  {journal} {\bibinfo  {journal} {Res. Policy}\ }\textbf {\bibinfo {volume}
  {28}},\ \bibinfo {pages} {1} (\bibinfo {year} {1999})}\BibitemShut {NoStop}%
\bibitem [{\citenamefont {Candia}\ \emph {et~al.}(2019)\citenamefont {Candia},
  \citenamefont {Jara-Figueroa}, \citenamefont {Rodriguez-Sickert},
  \citenamefont {Barab{\'a}si},\ and\ \citenamefont {Hidalgo}}]{can19}%
  \BibitemOpen
  \bibfield  {author} {\bibinfo {author} {\bibfnamefont {C.}~\bibnamefont
  {Candia}}, \bibinfo {author} {\bibfnamefont {C.}~\bibnamefont
  {Jara-Figueroa}}, \bibinfo {author} {\bibfnamefont {C.}~\bibnamefont
  {Rodriguez-Sickert}}, \bibinfo {author} {\bibfnamefont {A.-L.}\ \bibnamefont
  {Barab{\'a}si}}, \ and\ \bibinfo {author} {\bibfnamefont {C.~A.}\
  \bibnamefont {Hidalgo}},\ }\href {\doibase 10.1038/s41562-018-0474-5}
  {\bibfield  {journal} {\bibinfo  {journal} {Nat. Hum. Behav.}\ }\textbf
  {\bibinfo {volume} {65}},\ \bibinfo {pages} {82–91} (\bibinfo {year}
  {2019})}\BibitemShut {NoStop}%
\bibitem [{\citenamefont {Sorenson}\ \emph {et~al.}(2006)\citenamefont
  {Sorenson}, \citenamefont {Rivkin},\ and\ \citenamefont {Fleming}}]{sor06}%
  \BibitemOpen
  \bibfield  {author} {\bibinfo {author} {\bibfnamefont {O.}~\bibnamefont
  {Sorenson}}, \bibinfo {author} {\bibfnamefont {J.~W.}\ \bibnamefont
  {Rivkin}}, \ and\ \bibinfo {author} {\bibfnamefont {L.}~\bibnamefont
  {Fleming}},\ }\href {\doibase 10.1016/j.respol.2006.05.002} {\bibfield
  {journal} {\bibinfo  {journal} {Res. Policy}\ }\textbf {\bibinfo {volume}
  {35}},\ \bibinfo {pages} {994} (\bibinfo {year} {2006})}\BibitemShut
  {NoStop}%
\bibitem [{\citenamefont {Meyer}(2000)}]{mey00}%
  \BibitemOpen
  \bibfield  {author} {\bibinfo {author} {\bibfnamefont {M.}~\bibnamefont
  {Meyer}},\ }\href {\doibase 10.1023/A:1005613325648} {\bibfield  {journal}
  {\bibinfo  {journal} {Scientometrics}\ }\textbf {\bibinfo {volume} {49}},\
  \bibinfo {pages} {93} (\bibinfo {year} {2000})}\BibitemShut {NoStop}%
\bibitem [{\citenamefont {Clough}\ \emph {et~al.}(2015)\citenamefont {Clough},
  \citenamefont {Gollings}, \citenamefont {Loach},\ and\ \citenamefont
  {Evans}}]{clo15}%
  \BibitemOpen
  \bibfield  {author} {\bibinfo {author} {\bibfnamefont {J.~R.}\ \bibnamefont
  {Clough}}, \bibinfo {author} {\bibfnamefont {J.}~\bibnamefont {Gollings}},
  \bibinfo {author} {\bibfnamefont {T.~V.}\ \bibnamefont {Loach}}, \ and\
  \bibinfo {author} {\bibfnamefont {T.~S.}\ \bibnamefont {Evans}},\ }\href
  {\doibase 10.1093/comnet/cnu039} {\bibfield  {journal} {\bibinfo  {journal}
  {J. Complex Networks}\ }\textbf {\bibinfo {volume} {3}},\ \bibinfo {pages}
  {189} (\bibinfo {year} {2015})}\BibitemShut {NoStop}%
\bibitem [{\citenamefont {Caldarelli}\ \emph {et~al.}(2002)\citenamefont
  {Caldarelli}, \citenamefont {Capocci}, \citenamefont {De~Los~Rios},\ and\
  \citenamefont {Mu\~noz}}]{cal02}%
  \BibitemOpen
  \bibfield  {author} {\bibinfo {author} {\bibfnamefont {G.}~\bibnamefont
  {Caldarelli}}, \bibinfo {author} {\bibfnamefont {A.}~\bibnamefont {Capocci}},
  \bibinfo {author} {\bibfnamefont {P.}~\bibnamefont {De~Los~Rios}}, \ and\
  \bibinfo {author} {\bibfnamefont {M.~A.}\ \bibnamefont {Mu\~noz}},\ }\href
  {\doibase 10.1103/PhysRevLett.89.258702} {\bibfield  {journal} {\bibinfo
  {journal} {Phys. Rev. Lett.}\ }\textbf {\bibinfo {volume} {89}},\ \bibinfo
  {pages} {258702} (\bibinfo {year} {2002})}\BibitemShut {NoStop}%
\bibitem [{\citenamefont {Garlaschelli}\ and\ \citenamefont
  {Loffredo}(2004)}]{gar04}%
  \BibitemOpen
  \bibfield  {author} {\bibinfo {author} {\bibfnamefont {D.}~\bibnamefont
  {Garlaschelli}}\ and\ \bibinfo {author} {\bibfnamefont {M.~I.}\ \bibnamefont
  {Loffredo}},\ }\href {\doibase 10.1103/PhysRevLett.93.188701} {\bibfield
  {journal} {\bibinfo  {journal} {Phys. Rev. Lett.}\ }\textbf {\bibinfo
  {volume} {93}},\ \bibinfo {pages} {188701} (\bibinfo {year}
  {2004})}\BibitemShut {NoStop}%
\bibitem [{\citenamefont {Tacchella}\ \emph {et~al.}(2012)\citenamefont
  {Tacchella}, \citenamefont {Cristelli}, \citenamefont {Caldarelli},
  \citenamefont {Gabrielli},\ and\ \citenamefont {Pietronero}}]{tac12}%
  \BibitemOpen
  \bibfield  {author} {\bibinfo {author} {\bibfnamefont {A.}~\bibnamefont
  {Tacchella}}, \bibinfo {author} {\bibfnamefont {M.}~\bibnamefont
  {Cristelli}}, \bibinfo {author} {\bibfnamefont {G.}~\bibnamefont
  {Caldarelli}}, \bibinfo {author} {\bibfnamefont {A.}~\bibnamefont
  {Gabrielli}}, \ and\ \bibinfo {author} {\bibfnamefont {L.}~\bibnamefont
  {Pietronero}},\ }\href {\doibase 10.1038/srep00723} {\bibfield  {journal}
  {\bibinfo  {journal} {Sci. Rep.}\ }\textbf {\bibinfo {volume} {2}},\ \bibinfo
  {pages} {723} (\bibinfo {year} {2012})}\BibitemShut {NoStop}%
\bibitem [{\citenamefont {Mehta}\ \emph {et~al.}(2010)\citenamefont {Mehta},
  \citenamefont {Rysman},\ and\ \citenamefont {Simcoe}}]{meh10}%
  \BibitemOpen
  \bibfield  {author} {\bibinfo {author} {\bibfnamefont {A.}~\bibnamefont
  {Mehta}}, \bibinfo {author} {\bibfnamefont {M.}~\bibnamefont {Rysman}}, \
  and\ \bibinfo {author} {\bibfnamefont {T.}~\bibnamefont {Simcoe}},\ }\href
  {\doibase 10.1002/jae.1086} {\bibfield  {journal} {\bibinfo  {journal} {J.
  Appl. Econometrics}\ }\textbf {\bibinfo {volume} {25}},\ \bibinfo {pages}
  {1179} (\bibinfo {year} {2010})}\BibitemShut {NoStop}%
\end{thebibliography}
%

\end{document}